\documentclass[12pt]{iopart}
\usepackage{iopams}  
\usepackage{subfig,color}
\usepackage{graphicx}
\usepackage{cite}

\newcommand\rd{{\mathrm d}}
\def\bea{\begin{eqnarray}}
\def\eea{\end{eqnarray}}

\begin{document}

\title{The Jacobi last multiplier for linear partial difference equations}
\author{Decio Levi$^a$ and Miguel A. Rodr\'{\i}guez$^b$}

\address{$^a$Dipartimento di Ingegneria Elettronica, Universit\`a degli Studi Roma Tre and INFN Sezione di Roma Tre, Via della Vasca Navale 84, 00146 Roma, Italy  \\ $^b$Departamento de F\'{\i}sica Te\'{o}rica II, Facultad de F\'{\i}sicas, Universidad Complutense, 28040 Madrid, Spain}
\date{today}

\begin{abstract}
We present a discretization of the Jacobi last multiplier, with some applications to the computation of solutions of linear partial difference equations. 
\end{abstract}


\submitto{J. Phys. A: Math. Theor.}



\section{Introduction}
The Jacobi Last Multiplier (JLM) \cite{Bi18,Ja44} plays, in first order linear partial differential equations, a role  similar to the integrating factor in first order ordinary differential equations.  If one can guess a JLM, it is possible to find the general solution of the equation, or for equations with more than two variables to reduce the number of variables. However in the case of quasilinear first order partial differential equations we can always  integrate them going over to the characteristics. So the role of JLM is certainly not crucial as an integrating tool of PDEs, but it has an important role in many other applications. For example,  JLM has recently received a great deal of attention in  the theory of  $\lambda$-symmetries of differential equations \cite{NL11,LN12}. 

In this work, we present a first approach to an equivalent concept for difference equations. In the case of partial linear difference equations no integration technique equivalent to the use of the characteristics exists  \cite{Ch03,CF28} so the use of the JLM can be very proficuous.

Section 2 is devoted to a short review of JLM in first order partial differential equations, in particular the method to obtain the equation satisfied by a JLM. In section 3, the case of a difference equation in a two dimensional lattice is fully developed together with some examples. The conclusions  presented in Section 4 are devoted to a summary of the results obtained, showing the difficulties which appear when extending the method to a higher number of variables, and some future perspectives.


\section{The continuous Jacobi last multiplier: a review}\label{contin}

Let us consider a  first order linear partial differential equation:
\begin{equation}\label{veceq}
Xu=0,\quad X=\sum_{i=1}^N f^{(i)}\partial_{x^{(i)}}\, ,
\end{equation}
where $f^{(i)}$ are some smooth functions on the variables $x^{(i)}$. If $N-1$ functionally  independent solutions of this equation are known, $u^{(1)},\ldots,u^{(N-1)}$, we can write the following determinant
\begin{equation}\label{det}
\frac{\partial (u,u^{(1)},\ldots u^{(N-1)})}{\partial (x^{(1)},\ldots x^{(n)})}=\det\left(\begin{array}{ccc}\frac{\partial u}{\partial x^{(1)}}& \ldots & \frac{\partial u}{\partial x^{(n)}} \\[5pt]
\frac{\partial u^{(1)}}{\partial x^{(1)}}& \ldots & \frac{\partial u^{(1)}}{\partial x^{(n)}} \\
\vdots && \vdots \\
\frac{\partial u^{(N-1)}}{\partial x^{(1)}}& \ldots & \frac{\partial u^{(N-1)}}{\partial x^{(n)}}\end{array}\right) 
\end{equation}
for any function $u(x^{(1)},\ldots,x^{(n)})$.
The determinant (\ref{det}) is zero only if $u$ is a solution of equation (\ref{veceq}), as it can be easily shown since the elements of the matrix are the coefficients of the forms $\rd u^{(i)}$ and the functions $u,u^{(1)},\ldots,u^{(N-1)}$ are functionally dependent if $u$ is a solution of the differential equation.  Then the equations
\begin{equation}
Xu=0,\quad \frac{\partial (u,u^{(1)},\ldots u^{(N-1)})}{\partial (x^{(1)},\ldots x^{(n)})}=0
\end{equation}
have the same set of solutions and must be proportional as $N-1$ functionally independent solutions fix the coefficients of the linear first order PDE up to a global factor:
\begin{equation}\label{jac}
\frac{\partial (u,u^{(1)},\ldots u^{(N-1)})}{\partial (x^{(1)},\ldots x^{(n)})}=MXu.
\end{equation}
The multiplicative factor $M$ is called the Jacobi last multiplier.

If we expand the determinant (\ref{det}) using the first row, we find that equation (\ref{jac}) can be written as
\begin{equation}\label{last}
A^{(1)}\frac{\partial u}{\partial x^{(1)}}+\cdots +A^{(N)}\frac{\partial u}{\partial x^{(n)}}=M \sum_{i=1}^Nf^{(i)}\frac{\partial u}{\partial x^{(i)}},
\end{equation}
where $A^{(k)}$ are the corresponding minors of the first row of the matrix in (\ref{det}).  
Comparing  the coefficients of the derivatives of $u$ in (\ref{last}), we get
\begin{equation}\label{Adef}
A^{(k)}=Mf^{(k)},\quad k=1,\ldots,N.
\end{equation}

We can write now a differential equation satisfied by the functions  $A^{(k)}$
\begin{equation}\label{minor}
\fl 
(-1)^{k-1}\det\left(\begin{array}{cccccc}
\frac{\partial u^{(1)}}{\partial x^{(1)}}& \cdots & \frac{\partial u^{(1)}}{\partial x^{(k-1)}}& \frac{\partial u^{(1)}}{\partial x^{(k+1)}}&\cdots &\frac{\partial u^{(1)}}{\partial x^{(n)}} \\
\vdots && \vdots &\vdots && \vdots \\
\frac{\partial u^{(N-1)}}{\partial x^{(1)}}& \cdots & \frac{\partial u^{(N-1)}}{\partial x^{(k-1)}}& \frac{\partial u^{(N-1)}}{\partial x^{(k+1)}} & \cdots &  \frac{\partial u^{(N-1)}}{\partial x^{(n)}}
\end{array}\right)=A^{(k)},\quad k=1,\ldots,N.
\end{equation}
 This is a general relation arising from the particular form of the functions $A^{(k)}$, written as minors of the matrix (\ref{det}), and it does not depend on the fact that the functions $u^{(k)}$ are solutions of the linear homogeneous first order PDE (\ref{veceq}) with coefficients $f^{(k)}$. In fact, the equation for $A^{(k)}$  can be considered as a consistency condition for the system of equations in $u^{(j)}$ with $k=1,\ldots,N$, $j=1,\ldots,N-1$.

Let us  write (\ref{minor}) as 
\begin{equation} \label{minor1}
A^{(k)}=(-1)^{k-1}\det(B_1,\ldots,\widehat{B_k},\ldots, B_N),
\end{equation}
where by $B_i$ we denote the $i$-column of the matrix (\ref{det}) and the symbol $\widehat{B_k}$ means that the corresponding $k$th-column  of (\ref{det}) is removed.
Deriving (\ref{minor1}) with respect to $x^{(k)}$ and summing over $k$ we get:
\begin{equation}
\sum_{k=1}^N\frac{\partial A^{(k)}}{\partial x^{(k)}}=\sum_{i,k=1,i\neq k}^N(-1)^{k-1}\det(B_1,\ldots,\frac{\partial B_i}{\partial x^{(k)}},\ldots,\widehat{B_k},\ldots, B_N).
\end{equation}
Since
\begin{equation} \label{minor2}
\frac{\partial B_i}{\partial x^{(k)}}=\bigg(\frac{\partial^2 u^{(1)}}{\partial x^{(i)}\partial x^{(k)}},\ldots,\frac{\partial^2u^{(N-1)}}{\partial x^{(i)}\partial x^{(k)}}\bigg)^T=\frac{\partial B_k}{\partial x^{(i)}},
\end{equation}
we have
\begin{eqnarray}\fl
\det(B_1,\ldots,\frac{\partial B_i}{\partial x^{(k)}},\ldots,\widehat{B_k},\ldots, B_N)=\det(B_1,\ldots,\frac{\partial B_k}{\partial x^{(i)}},\ldots,\widehat{B_k},\ldots, B_N)\nonumber \\ =(-1)^{k-i-1}\det(B_1,\ldots,\widehat{B_i},\ldots,\frac{\partial B_k}{\partial x^{(i)}},\ldots, B_N).
\end{eqnarray}
Finally, since $(-1)^k(-1)^{k-i-1}=-(-1)^{i}$ the sum of all terms in (\ref{minor2}) is zero and consequently we get the equation:
\begin{equation}\label{ecA}
\frac{\partial A^{(1)}}{\partial x^{(1)}}+\cdots +\frac{\partial A^{(N)}}{\partial x^{(n)}}=0.
\end{equation}
In three dimensions, this expression is the classical formula of vector calculus stating that the divergence of the cross product of two gradient vectors is zero. In an arbitrary dimension it can be written using exterior products and differential forms (see for instance \cite{GG09}).

We can derive an equation for $M$, differentiating (\ref{Adef}) with respect to $x^{(k)}$ and summing over all $k$ between 1 and $N$. We get
\begin{equation}
\sum_{k=1}^N\frac{\partial A^{(k)}}{\partial x^{(k)}}=M\sum_{k=1}^N \frac{\partial f^{(k)}}{\partial x^{(k)}}+\sum_{k=1}^N f^{(k)}\frac{\partial M}{\partial x^{(k)}}
\end{equation}
and consequently, taking into account  (\ref{ecA}), we obtain:
\begin{equation}\label{lastM}
\sum_{k=1}^{N}f^{(k)}\frac{\partial \log M}{\partial x^{(k)}}+\sum_{k=1}^{N}\frac{\partial f^{(k)}}{\partial x^{(k)}} =0.
\end{equation}
This equation depends only on  the differential equation (\ref{veceq}) and thus $M$ does not depend on any particular solution.  From a practical point of view, equation (\ref{lastM}) is an inhomogeneous version of the original equation (\ref {veceq}). However, as in the method of integrating factors, if we know a particular solution of (\ref{lastM}), we could use it to compute a  solution of (\ref{veceq}). This is exploited in the examples presented in the next subsection where we consider a few examples and  we compare the results obtained by the use of JLM with those obtained through other methods of integration of the  linear first order partial differential equations.
 

\subsection{ Examples} 

As a simple illustration of the method, let us consider the following examples of partial differential equations in two and three independent variables.
\subsubsection{Two independent variables}
\begin{equation}
yu_x+xu_y=0.
\end{equation}

From (\ref{lastM}) the equation for $M$ is:
\begin{equation}
y\partial_x\log M+x\partial_y\log M=0.
\end{equation}
An obvious solution of this equation is $M=1$. Then, 
\begin{equation}
A^{(1)}=Mf_1=y,\quad A^{(2)}=Mf_2=x
\end{equation}
and (\ref{Adef}) reduces to the compatible overdetermined system of equations:
\begin{equation}
 u_x=-x,\quad u_y=y.
\end{equation}
Solving this system we  find a non trivial solution of the original partial differential equation
\begin{equation}
u(x,y)=\frac12(y^2-x^2).
\end{equation}
The general solution can be obtained by computing the characteristic variable $\xi=y^2-x^2$ and is given by:
\begin{equation}
u(x,y)=F(y^2-x^2),
\end{equation}
where $F$ is an arbitrary function of its argument defined by the initial conditions or from the boundary values. 

\subsubsection{Three independent variables}

\begin{equation} \label{e4}
x(x+y)u_x-y(x+y)u_y+z(x-y)u_z=0.
\end{equation}
The equation satisfied by a Jacobi last multiplier $M$ is
\begin{equation}\fl 
x(x+y)\frac{\partial}{\partial x}\log M-y(x+y)\frac{\partial}{\partial y}\log M+z(x-y)\frac{\partial}{\partial z}\log M+2(x-y)=0.
\end{equation}
Looking for a particular solution of this equation, for instance  $M=M(z)$, we find $M=\frac{1}{z^2}$, and the system of equations we have to solve is (with $u^{(1)}\equiv u$, $u^{(2)}\equiv v$):
\begin{equation}\fl
u_yv_z-u_zv_y=\frac{x(x+y)}{z^2},\quad u_zv_x-u_xv_z=-\frac{y(x+y)}{z^2},\quad u_xv_y-u_yv_x=\frac{x-y}{z}. \label{e4a}
\end{equation}
Given  a solution $u$  of (\ref{e4}), (\ref{e4a}) is an overdetermined system for $v$. For instance, we can take $u(x,y,z)=xy$ and the new  solution $v$ is obtained  from the following overdetermined system of equations 
\begin{equation} \label{e4b}
v_z=\frac{x+y}{z^2},\quad  yv_y-xv_x=\frac{x-y}{z}.
\end{equation}
A solution of (\ref{e4b}) is
\begin{equation}
v=-\frac{x+y}{z}.
\end{equation}
From the method of characteristics we find that any solution of (\ref{e4}) is a function of the two particular solutions we have found
\begin{equation}
u(x,y,z)=F\left(xy,\frac{x+y}{z}\right).
\end{equation}


\section{Difference equations}

A difference equations for one dependent variable $u$ is is a relation between the  function in various points of a lattice. If the lattice is $r$ dimensional it can be put in correspondence with the points of an $r$--dimensional space. 

An ordinary difference equation (O$\Delta$E) is a difference equation on a one dimensional lattice. In this case  the lattice is given by an ordered sequence of points on a line characterized by their relative distance (see Fig.1). If $x_i$ and $x_{i+1}$ are two subsequent points, their distance $|x_{i+1}-x_i|$ will be $h_i$. We can then introduce a $x$--shift operator $T_x$ such that $T_x x_i=x_{i+1}$ and in term of it we can construct delta operators which in the continuous limit, when $h_i\rightarrow 0$, go over to the derivative. An example of such delta operator is given by  the right shifted discrete derivative
\bea \label{d1} 
\Delta_x u(x_i)=\frac{u(x_{i+1}) -u(x_i)}{x_{i+1} -x_i} = \frac{(T_x-1) u(x_i)}{h_i}.
\eea
In some instances it may be more convenient to introduce symmetric delta operators \cite{ltw} as, for example,
\bea \label{d2} 
\Delta_x^s u(x_i)=\frac{u(x_{i+1}) -u(x_{i-1})}{x_{i+1} -x_{i-1}} = \frac{(T_x-T_x^{-1}) u(x_i)}{h_i+h_{i-1}}.
\eea
An O$\Delta$E of order $n$, i.e. involving $n+1$ points of the lattice, can thus be written as 
\bea \label{d3} 
\mathcal E(x_i, u_i, T_x u_i, T_x^2 u_i, \cdots T_x^n u_i)=0,
\eea
or, equivalently, in term of the operator delta as
\bea \label{d4} 
\mathcal F(x_i, u_i, \Delta_x u_i, \Delta_x^2 u_i, \cdots \Delta_x^n u_i)=0.
\eea
However the equation (\ref{d3}) (or (\ref{d4})) is not completely defined unless we specify the values of the distance between the various lattice points $h_j$, $j=i,...,i+n$ involved in the equation. This implies that an O$\Delta$E will be defined only if we attach to it a second equation which defines  the lattice. The set of these two equations is  called a {\it Difference scheme}. 

In many instances, when the equation comes from some physical problem, the lattice is a priori given, as for example, when all the points are equidistant so that  $h_i=h$. In this case the lattice equation is trivial $x_{i+1}-x_i=h$.  However there may be situations, as, for example,  discretizing a continuous differential equation to solve it on the computer, when we want to take advantage of the freedom of the lattice to preserve in the discretization other properties of the continuous system like its symmetries \cite{lw2006}. In such a situation the lattice may be defined by a non trivial equation maybe also depending on the dependent variable so as to have a denser grid when the solution varies rapidly. 

\begin{figure}
  \centering
  \subfloat[the index line]{\label{fig:plane}\includegraphics[width=0.3\textwidth]{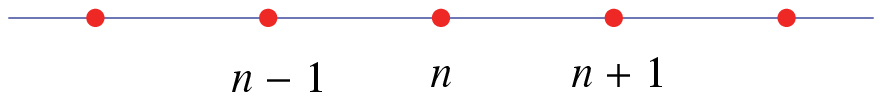}} \qquad 
  \subfloat[the $x$--line]{\label{fig:xy}\includegraphics[width=0.3\textwidth]{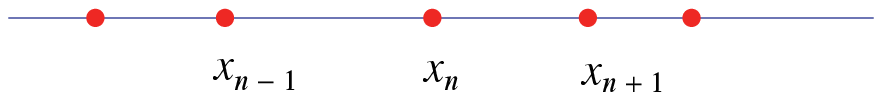}}
  ~ 
    \caption{ $1$--dimensional lattice grids}
  \label{fig:grids}
\end{figure}\begin{figure}
  \centering
  \subfloat[the index plane]{\label{fig:plane}\includegraphics[width=0.3\textwidth]{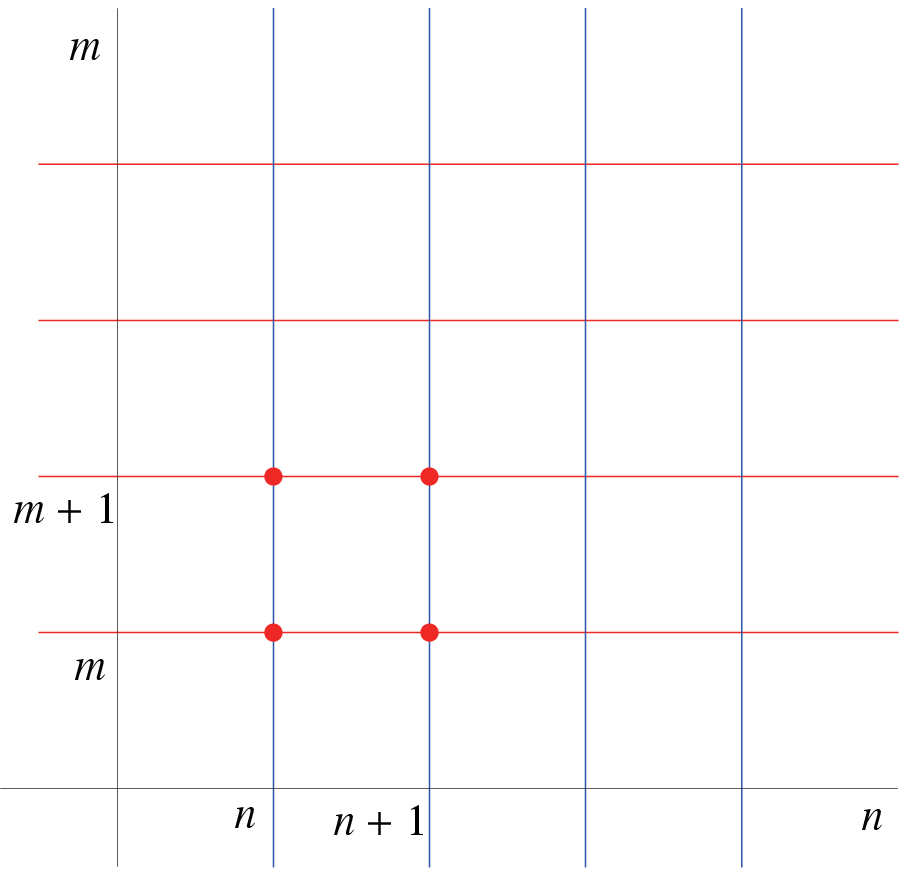}} \qquad 
  \subfloat[the $x,y$--plane]{\label{fig:xy}\includegraphics[width=0.3\textwidth]{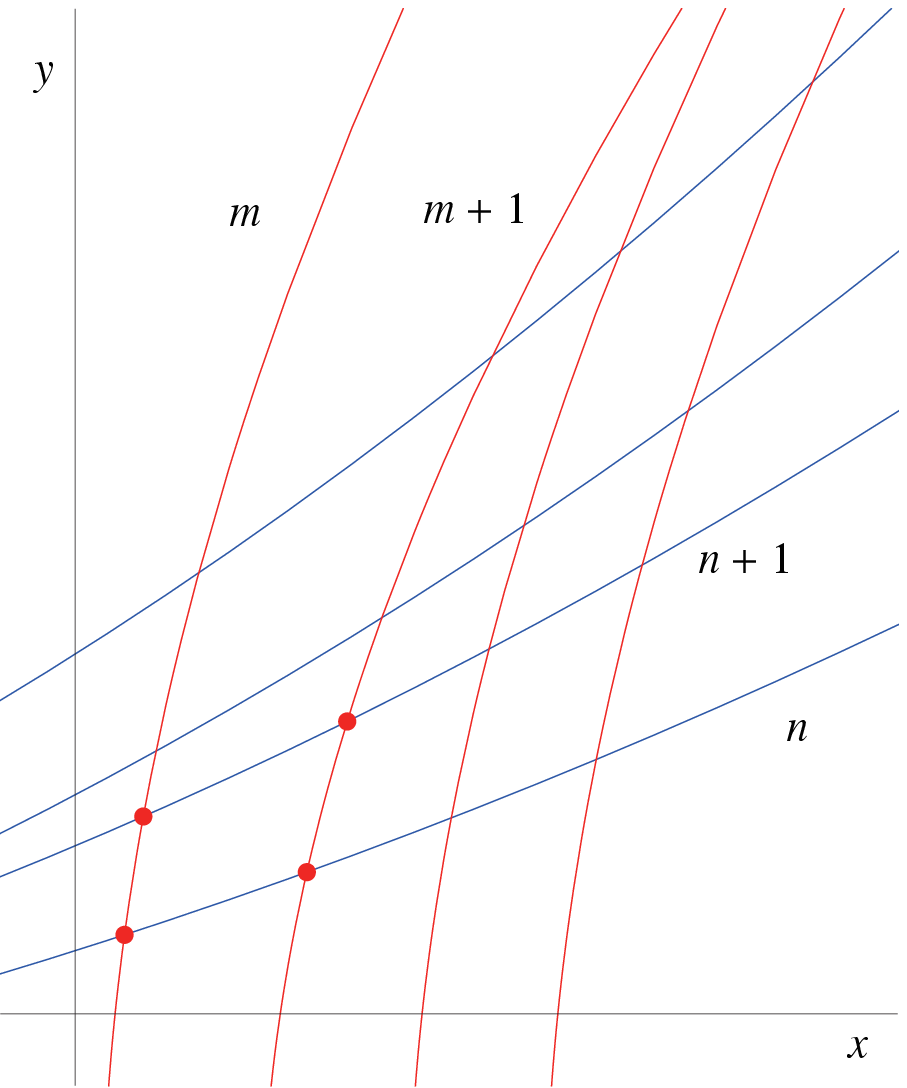}}
  ~ 
    \caption{ $2$--dimensional lattice grids}
  \label{fig:grids2}
\end{figure}

In the case of O$\Delta$E's there is at least one  natural parametrization of the differences which in the continuos limit go to the corresponding derivatives and which simplifies the discretization procedure \cite{ltw2010}. Such parametrization gives  a one-to-one transformation between the lattice differences, discrete approximations of the derivatives, and the lattice points. 

A similar situation exists also in the case of partial difference equations (P$\Delta$E's), however in this case the definition of the lattice must be given by compatible equations as the independent variables depend on several indices (see Fig.2 for the $2$--dimensional case where $x_{n,m}$ and $y_{n,m}$, depend on two indices).  In general the difference scheme will be given apart from the  P$\Delta$E, by a set of equations which define the lattice and depend on the number of independent variables and on the problem we are solving \cite{ltw}. However in this case, up to now, no natural parametrization exists which in the continuous limit goes to the corresponding derivatives and which simplifies the discretization procedure. Work on this is in progress \cite{rv2012,LR12}. 

The solution of linear O$\Delta$E's follows the standard technique of solving ordinary differential equations. The solution is given by a linear combination with arbitrary coefficients of powers of the independent variable  and the exponents are defined by a characteristic polynomial.  In the case of linear P$\Delta$E's the situation is more complicate (as is also the case for partial differential equations). As one can read in \cite{Ch03} {\it The method of trial and error is still one of the basic methods for obtaining explicit solutions}. If the P$\Delta$E  has constant coefficients then a few techniques can be found in Jordan book \cite{j1950}, such as Laplace method of generating functions, or the method of Fourier, Lagrange and Ellis \cite{E1844}. If the P$\Delta$E does not depend explicitly on one of the two independent variables then  Boole symbolic method can be applied \cite{b}.  

Consequently, it seems particularly important to extend the last Jacobi multiplier technique to the case of linear P$\Delta$E's as this will provide solutions also in the case of  non constant coefficient P$\Delta$E's. In the following, for the sake of simplicity, we will consider the case of an  orthogonal lattice,  where the independent variables $x_{n,m}$ and $y_{n,m}$ can be written in term of just one index, i.e.  $x_n$ and $y_m$, and we will just concentrate on the solution of the difference equation. 

\subsection{Jacobi last multiplier on a  two dimensional lattice}
 Let us write an equivalent discrete expression defined on 4 lattice points of the linear first order partial differential equation (\ref{veceq}) . We could write this expression in terms of the shift operators but we will  use the difference operators (\ref{d1}) to follow closely the continuous case and to limit the number of points involved.
Let us  consider a two dimensional orthogonal lattice. In such a case (\ref{veceq}) reads:
\begin{eqnarray}\label{dif}
\fl \qquad f^{(1)}_{n,m}(x_{n},x_{n+1},y_m,y_{m+1})\Delta_x u_{n,m}  +f^{(2)}_{n,m}(x_{n},x_{n+1},y_m,y_{m+1})\Delta_y u_{n,m}=0.
\end{eqnarray}
As in the continuous case, 
if $u^{(1)}_{n,m}$ is a solution of  (\ref{dif}), the   $2\times 2$ matrix  
\begin{equation} \label{matrix}
\frac{\Delta(u_{n,m},u^{(1)}_{n,m})}{\Delta(x_{n},y_m)}=
\left(\begin{array}{cc}
\Delta_x u_{n,m} & \Delta_y u_{n,m} \\[5pt]
\Delta_x u^{(1)}_{n,m} & \Delta_y u^{(1)}_{n,m}
\end{array}\right)\end{equation}
has a determinant equal to zero if and only if the function $u_{n,m}$ is also a solution of the difference equation (\ref{dif}). The minors of the first row  of (\ref{matrix}) are
\begin{equation}\label{ff1}
A^{(1)}_{n,m}=\Delta_x u^{(1)}_{n,m},\quad A^{(2)}_{n,m}=-\Delta_y u^{(1)}_{n,m}
\end{equation}
and it is trivial to check by direct computation that $A^{(1)}$ and $A^{(2)}$ satisfy the  equation
\begin{equation}
\Delta_x A^{(1)}_{n,m}+\Delta_y A^{(2)}_{n,m}=0.
\end{equation}
Then, as in the continuous case, there exists a function $M_{n,m}$, the Jacobi last multiplier, such that, 
\begin{equation}\label{f1}
A^{(1)}_{n,m}=M_{n,m}f^{(1)}_{n,m},\quad A^{(2)}_{n,m}=M_{n,m} f^{(2)}_{n,m}.
\end{equation}
Consequently,  the Jacobi last multiplier $M_{n,m}$ satisfies a difference equation, which is the discrete analog of the differential one  (\ref{lastM})
\begin{equation}\label{f3a}
\frac{1}{M_{n,m}}\bigg(\Delta_x M_{n,m} f^{(1)}_{n+1,m}+\Delta_y M_{n,m} f^{(2)}_{n,m+1}\bigg) +\Delta_x f^{(1)}_{n,m}+\Delta_y f^{(2)}_{n,m}=0.
\end{equation}
Given any particular, even trivial, solution $M_{n,m}$ of the P$\Delta$E (\ref{f3a}), the solution of the overdetermined system (\ref{f1}) provides a solution of (\ref{dif}). We consider now a few examples of the calculus of the solution of linear difference equations using the JLM.

\subsection{Examples}


\subsubsection{First example.}

We consider the equation:
 \begin{equation} \label{pippo1}
 y_m\Delta_x u_{n,m} +x_n\Delta_y u_{n,m}=0.
 \end{equation}
The equation for the Jacobi last multiplier  $M_{n,m}$ is the same as that for $u_{n,m}$:
 \begin{equation}
 y_m\Delta_x M_{n,m} +x_n\Delta_y M_{n,m}=0,
 \end{equation}
and a particular solution is obviously $M_{n,m}=1$ which gives, taking into account (\ref{f1}), the following system of equations for $u_{n,m}$:
 \begin{equation}\label{f2}
\Delta_x u_{n,m}=-x_n,\quad \Delta_y u_{n,m}=y_m.
 \end{equation}
To solve this sytem of difference equations we need to specify the lattice.  If we consider a uniform lattice in both variables i.e.:
\begin{equation}
x_{n+1}-x_n=\delta_1,\quad y_{m+1}-y_m=\delta_2,
\end{equation}
so that $x_n=\delta_1 n+x_0$ and $y_m= \delta_2 m + y_0$, where $x_0$ and $y_0$ are arbitrary initial points, 
 the system (\ref{f2}) reduces  to a system of O$\Delta$E's, one for each direction:
\begin{equation} \label{f2a}
u_{n+1,m}=u_{n,m}-n\delta_1^2-\delta_1 x_0,\quad  u_{n,m+1}=u_{n,m}+ m\delta_2^2+\delta_2y_0.
\end{equation}
Using the well know procedures for solving  O$\Delta$E's \cite{j1950} we get a particular solution of (\ref{pippo1}), depending on three arbitrary constants, i.e.:
\begin{equation}
 u_{n,m}=u_{0,0}-\frac12(x_n+x_0)(x_n-x_0-\delta_1)+\frac12(y_m+y_0)(y_m-y_0-\delta_2).
\end{equation}


\subsubsection{Second example.}

We choose the equation:
\begin{equation}\label{eq2}
 y_m x_{n+1}\Delta_x u_{n,m} +x_n y_{m+1}\Delta_y u_{n,m}=0,
\end{equation}
which corresponds to  (\ref{dif}) with
 \begin{equation}
 f_{n,m}^{(1)}= y_m x_{n+1},\qquad f^{(2)}_{n,m}=x_n y_{m+1}.
\end{equation}
The equation for the Jacobi last multiplier can be written as:
\begin{eqnarray}\fl\label{f3}
\frac{y_m}{x_{n+1}- x_{n}} \Big \{\frac{M_{n+1,m}}{M_{n,m}}  x_{n+2} -  x_{n+1} \Big \} +\frac{x_{n}}{y_{m+1}- y_{m}} \Big \{\frac{M_{n,m+1}}{M_{n,m}} y_{m+2}  - y_{m+1}  \Big \} =0.
\end{eqnarray}
A particular solution of the equation (\ref{f3})  can be obtained requiring that both its curly brackets be identically zero. In such a case we get  a particular solution 
\begin{equation}
M_{n,m}= \frac{\alpha}{y_{m+1}x_{n+1}},
\end{equation}
where $\alpha$ is an abitrary constant. If we introduce this result in (\ref{f1})  with $A^{(1)}_{n,m}$ and $A^{(2)}_{n,m}$ given by (\ref{ff1}), we get the following  system of compatible equations 
\begin{equation}\eqalign{ \label{eq2a}
u_{n+1,m} - u_{n,m} = - \alpha x_n \left ( 1 - \frac{x_n}{x_{n+1}} \right), \\ 
u_{n,m+1} - u_{n,m} = \quad \alpha y_m \left( 1 - \frac{y_m}{y_{m+1}} \right).}
\end{equation}

As in the previous example, we need the lattice equations to solve equations (\ref{eq2a}). Using again a uniform lattice in each variable, we get 
\bea
x_n=x_0+h_xn,\quad y_m=y_0+h_y m, \\ \eqalign{
u_{n+1,m} - u_{n,m} = - \alpha h_x\left (1- \frac{h_x}{x_0+(n+1)h_x}\right),  \\
u_{n,m+1} - u_{n,m} = \quad \alpha h_y\left (1-  \frac{h_y}{y_0+(m+1)h_y}\right).\label{syst} }
\eea

To solve the first equation we define,  
\begin{equation}
v_{n,m}=\frac{1}{\alpha h_x}u_{n,m}+n
\end{equation}
and the equation satisfied by $v_{n,m}$ is:
\begin{equation}
v_{n+1,m}= v_{n,m}+\frac{1}{1+\frac{x_0}{h_x}+ n},
\end{equation}
which is the recursion equation for the Euler digamma function $\psi$.
Then
\begin{equation}\label{sol}
 v_{n,m}=a_m+\psi\left(n+\frac{x_0}{h_x}+1\right).
\end{equation}

Substituting in the second equation in (\ref{syst}) we easily obtain an equation fo $a_m$:
\begin{equation}
a_{m+1}=a_m -\frac{h_y}{h_x}\left(1-\frac{ h_y}{y_0 +(m+1)h_y}\right).
\end{equation}
Solving this equation as in  (\ref{syst}),
\begin{equation} 
a_{m}= \frac{h_y}{h_x} \left(c+\psi\left(m+\frac{y_0}{h_y}+1\right)-m\right),
\end{equation}
where $c$ is a constant.
Then, the complete solution is
\begin{eqnarray}
 u_{n,m}= &u_{0,0}+\alpha \Bigg[x_0 -y_0  - h_x \psi\left(1+\frac{x_0}{h_x}\right)+  h_y \psi\left(1+\frac{y_0}{h_y}\right)-\nonumber \\  &-  x_n+  y_m +   h_x  \psi\left(1 +\frac{x_n}{h_x}\right)-  h_y  \psi\left(1+\frac{y_m}{h_y}\right)\Bigg].
\end{eqnarray}


\section{Conclusions}

In this paper we have extended the results presented by Jacobi in 1844 to get solutions of linear partial differential equations to the case of partial difference equations in two independent variables.   

If we consider an $N$-dimensional lattice of independent coordinates, that is the lattice coordinates, $x^{(i)}$,  depend only on one index $n_i$, $i=1,\ldots, N$, it is easy to see that we get into trouble as the minors are nonlinear functions and the difference operator, in contrast with the differential one, does not satisfy Leibniz rule. In fact denoting   by $u_{n_1,\ldots,n_N}$  the value of $u$ at the point $(x^{(1)}_{n_1},\ldots,x^{(N)}_{n_N})$ and using the following notation:
\begin{equation}
 \boldsymbol{n}=(n_1,\ldots,n_N),\quad \boldsymbol{x}_{\boldsymbol{n}}=(x^{(1)}_{n_1},\ldots, x^{(N)}_{n_N}),\quad  \boldsymbol{\epsilon}_i=(0,\ldots,1(i),\ldots,0),
\end{equation}
 the discrete derivatives reads:
\begin{equation}
\Delta_{i} u_{\boldsymbol{n}}=\frac{u_{\boldsymbol{n}+\boldsymbol{\epsilon}_i}-u_{\boldsymbol{n}}}{x^{(i)}_{n_i+1}-x^{(i)}_{n_i}}.
\end{equation}

The difference equation is: 
\begin{equation}\label{three}
\sum_{i=1}^{N} f^{(i)}_{\boldsymbol{n}}\Delta_{i} u_{\boldsymbol{n}}=0,
\end{equation}
where $f^{(i)}_{\boldsymbol{n}}$ are some functions depending on a finite number of points in the lattice. If we know $N-1$ particular solutions of the equation, $u^{(1)}_{\boldsymbol{n}},\ldots,u^{(N-1)}_{\boldsymbol{n}}$, we can construct the matrix
\begin{equation}
\frac{\Delta(u_{\boldsymbol{n}},u^{(1)}_{\boldsymbol{n}},\ldots,u^{(N-1)}_{\boldsymbol{n}})}{\Delta(\boldsymbol{x}_{\boldsymbol{n}})}=
\left(\begin{array}{ccc}
\Delta_{1} u_{\boldsymbol{n}} & \cdots & \Delta_{N} u_{\boldsymbol{n}}
\\ 
\Delta_{1} u^{(1)}_{\boldsymbol{n}} & \cdots &\Delta_{N} u^{(1)}_{\boldsymbol{n}}
\\
\vdots & & \vdots\\
\Delta_{1} u^{(N-1)}_{\boldsymbol{n}} & \cdots & \Delta_{N} u^{(N-1)}_{\boldsymbol{n}}
\end{array}\right)
\end{equation}
and define the minors $A^{(k)}_{\boldsymbol{n}}$, $k=1,\ldots ,N$, corresponding to the first line of the matrix above (the column $k$ is removed):  
\begin{equation}\label{Adiscrete}
A^{(k)}_{\boldsymbol{n}}=(-1)^{k-1}\det
\left(\begin{array}{ccccc}
\Delta_{1} u^{(1)}_{\boldsymbol{n}} & \cdots & \hat{k} & \cdots & \Delta_{N} u^{(1)}_{\boldsymbol{n}}
\\
\vdots & & \vdots& & \vdots\\
\Delta_{1} u^{(N-1)}_{\boldsymbol{n}}& \cdots & \hat{k} & \cdots & \Delta_{N} u^{(N-1)}_{\boldsymbol{n}}
\end{array}\right).
\end{equation}
As in the continuous case (see Section \ref{contin}), the $N-1$ solutions fix the coefficients of the difference equation (\ref{three}) up to a factor. Then, using Cramer's rule, we get
\begin{equation}\label{eqM}
A^{(k)}_{\boldsymbol{n}}=M_{\boldsymbol{n}} f^{(k)}_{\boldsymbol{n}},\quad k=1,\ldots, N.
\end{equation}

To find the compatibility relation we will closely follow the argument we used in the continuous case. Then, writing
\begin{equation}
A^{(k)}_{\boldsymbol{n}}=(-1)^{k-1}\det(B_1,\ldots,\widehat{B_k},\ldots, B_N).
\end{equation}
where  $B_i$ is the $i$-the column of (\ref{Adiscrete}) and $\hat{B}_k$ means that the $k$-th column is absent,  we get that the sum
\begin{equation}\label{sumdisc}
\sum_{i,k=1,i\neq k}^N(-1)^{k-1}\det(B_1,\ldots,\Delta_{k} B_i,\ldots,\widehat{B_k},\ldots, B_N)
\end{equation}
is equal to zero. In fact
\begin{equation}
\Delta_{k} B_i=\bigg(\Delta_{k} \Delta_{i} u_{\boldsymbol{n}}^{(1)},\ldots,\Delta_{k} \Delta_{i} u_{\boldsymbol{n}}^{(N-1)}\bigg)^T=\Delta_{i} B_k,
\end{equation}
since $\Delta_{k} \Delta_{i}=\Delta_{i} \Delta_{k}$ as the independent variables commute. Then
\begin{eqnarray}\fl
\det(B_1,\ldots,\Delta_{k} B_i,\ldots,\widehat{B_k},\ldots, B_N)=\det(B_1,\ldots,\Delta_{i} B_k,\ldots,\widehat{B_k},\ldots, B_N)\nonumber \\  =(-1)^{k-i-1}\det(B_1,\ldots,\widehat{B_i},\ldots,\Delta_{i} B_k,\ldots, B_N)
\end{eqnarray}
and the sum (\ref{sumdisc})  is zero.
This is exactly the same equation we found in the continuous case. However, since Leibniz rule does not apply in the case of difference operators, the expression (\ref{sumdisc}) is not equal to $\sum_{k=1}^N\Delta_{k} A^{(k)}_{\boldsymbol{n}}$,  as the difference operator  does not follow the same rules as the differential operator when it is applied to a determinant. There are some additional terms, as $\Delta(fg)=f\Delta g+g\Delta f+h(\Delta f)(\Delta g)$, whose consequences have to be analyzed. They will be the content of a future work.


\section*{Acknowledgments}
DL  has been partly supported by the Italian Ministry of Education and Research, 2010 PRIN ``Continuous and discrete nonlinear integrable evolutions: from water waves to symplectic maps". MAR was supported by the Spanish Ministry of Science and Innovation under project  FIS2011-22566 and Universidad Complutense under project GR35/10-A910556. This article was completed during a stay of MAR at the Dipartimento di Ingegneria Elettronica of Universit\`a degli Studi Roma Tre (Italy). MAR would like to thank Fundaci\'on Caja Madrid of Spain for the financial support for this stay and the Universit\`a degli Studi Roma Tre for its hospitality.



\end{document}